# Preprint version of the manuscript

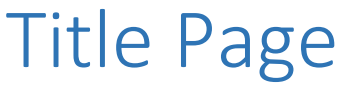

**Title:**

**Two for One – Combined Morphologic and Quantitative Knee Joint MRI Using a Versatile Turbo Spin-Echo Platform**

**Authors**

Teresa Lemainque[1], Nicola Pridoehl[1], Marc Huppertz[1], Manuel Post[1], Can Yüksel[1], Karl Ludger Radke[2], Shuo Zhang[3], Masami Yoneyama[4], Andreas Prescher[5], Christiane Kuhl[1], Daniel Truhn[1], Sven Nebelung[1]

**Institutions**

[1]Department of Diagnostic and Interventional Radiology, Medical Faculty, RWTH Aachen University, Aachen, Germany

[2]Department of Diagnostic and Interventional Radiology, University Hospital Düsseldorf, University Dusseldorf, Düsseldorf, Germany

[3]Philips GmbH Market DACH, Hamburg, Germany

[4]Philips Japan, Tokyo, Japan

[5]Institute of Molecular and Cellular Anatomy, Medical Faculty, RWTH Aachen University, Aachen, Germany

**Corresponding author**

Teresa Lemainque, M.Sc., Ph.D.
Department of Diagnostic and Interventional Radiology
Medical Faculty
RWTH Aachen University
Pauwelsstraße 30
52074 Aachen
Germany
E-Mail: tlemainque@ukaachen.de
Tel: +49 241 80 37843

**Funding declaration:**

This work was supported by the START program of the Faculty of Medicine of the RWTH Aachen University (No. 692316). DT is supported by the European Union's Horizon Europe programme (ODELIA, 101057091), by grants from the Deutsche Forschungsgemeinschaft (DFG) (TR 1700/7-1), and the German Federal Ministry of Education and Research (SWAG, 01KD2215A; TRANSFORM LIVER, 031L0312A). SN is funded by grants from the Deutsche Forschungsgemeinschaft (DFG) (NE 2136/3-1).

Word count: currently 3453 words

# Abstract

**Introduction**

Quantitative MRI techniques such as $T_2$ and $T_{1\rho}$ mapping are beneficial in evaluating knee joint pathologies; however, long acquisition times limit their clinical adoption. MIXTURE (**M**ulti-**I**nterleaved **X**-prepared **T**urbo-Spin Echo with Int**U**itive **RE**laxometry) provides a versatile turbo spin-echo (TSE) sequence platform for simultaneous morphologic and quantitative joint imaging yet lacks comparative evaluation in basic and translational research contexts.

**Methods**

Two MIXTURE sequences were designed along clinical requirements: (i) 'MIX1', combining proton density (PD)-weighted fat-saturated (FS) images and quantitative T2 mapping (acquisition time: 4:59 min), and (ii) 'MIX2', combining T1-weighted images with quantitative T1ρ mapping (6:38 min). MIXTURE sequences and their reference 2D and 3D TSE counterparts were acquired from ten human cadaveric knee joints using a clinical 3T MRI scanner and knee coil. Contrast, contrast-to-noise ratios, and coefficients of variation were comparatively evaluated using parametric tests. Clinical radiologists (n=3) assessed diagnostic quality as a function of sequence and anatomic structure using 5-point Likert scales and ordinal regression. The significance level was set to $\alpha$=0.01.

**Results**

MIX1 and MIX2 had at least equal diagnostic quality compared to the 2D and 3D TSE sequences of the same image weighting. Contrast, contrast-to-noise ratios, and coefficients of variation were largely similar for the PD-weighted FS and T1-weighted images.

**Discussion**

In clinically feasible scan times, the MIXTURE sequence platform yields (i) morphologic images of diagnostic quality and adjustable TSE-based contrasts and (ii) quantitative parameter mapping with additional insights on soft tissue composition and ultrastructure.

## Introduction

Osteoarthritis (OA) is a chronic degenerative joint disease with high prevalence and economic burden [1]. Cartilage degeneration is the structural hallmark of OA, while the surrounding tissues, i.e., menisci, bones, ligaments/tendons, and synovial tissue, are affected, too [2]. Magnetic resonance imaging (MRI) is the superordinate imaging modality to assess OA because it provides excellent soft tissue contrast [3].

The relaxation times T2 and T1ρ have been explored as surrogate markers of tissue composition, ultrastructure, and functionality of cartilage [2] [4] [5]. These parameters are sensitive to multiple constituents of cartilage at a time, but none is specific to a singular component [6]. For example, T2 was reported to correlate positively with water content and the orientation of collagen fibres, and negatively with the collagen and proteoglycan content. Likewise, assessing these parameters can help detect early OA, i.e., when the damage may still be reversible [7] [8]. T2 mapping, when added to the clinical routine protocol, increased the sensitivity of detecting early OA [9] [10]. However, the clinical adoption of T2 and T1ρ mapping techniques is -among other factors- impeded by their long acquisition times, which may take 10 min or more to cover the whole joint [11].

Combined MRI sequences aim to solve this problem by providing (diagnostically useful) morphologic images alongside quantitative parameter maps in clinically feasible scan times. In knee joint imaging, the quantitative double-echo-steady-state (qDESS) sequence provides T2 maps based on two gradient-echo images and takes 5 min to cover the whole joint [12]. The morphologic qDESS images enabled the comprehensive joint diagnosis with similar sensitivity, specificity, and accuracy compared to the conventional images [13]. However, the gradient-echo image contrast differs substantially from the turbo spin echo (TSE) image contrast predominantly used clinically. Hence, **M**ulti-**I**nterleaved **X**-prepared **T**urbo-Spin Echo with Int**U**itive **Re**laxometry (MIXTURE) has been introduced as a sequence platform to combine two or more TSE-based image contrasts with simultaneous T2 or T1ρ mapping. First preliminary studies [14] [15] [16] provide showed that whole-joint T2 and T1ρ mapping at isotropic resolution may be combined with morphologic image contrasts, e.g., proton density (PD)-weighted (-w) and $T_2$-w fat-saturated (FS) images.

Nevertheless, beyond technical feasibility, certain aspects related to the sequence platform's clinical utilization, such as diagnostic quality and time demand, remain to be elucidated. Consequently, we aimed to (i) develop and refine MIXTURE sequence variants to match our institution's standard knee protocol and (ii) evaluate these variants versus the corresponding 2D and 3D TSE reference sequences in basic and translational research contexts. We hypothesized that MIXTURE sequences provide (i) morphologic images with characteristic TSE-based contrasts and sufficient diagnostic quality and (ii) quantitative maps simultaneously with only moderately increased time demand.

# Methods

## Study Design and Sample Size Estimation

The local Institutional Review Board approved this prospective in-situ imaging study on human cadaveric knee joint specimens (Ethical Committee, RWTH Aachen University, EK180/16). The body donors had given written informed consent prior to the study. Assuming an inter-sequence difference in diagnostic quality scores of 0.5, an inter-specimen standard deviation (SD) of 0.4, a power of 0.8, and a significance level of 0.01, the minimum sample size was determined as eight. Ten fresh-frozen and non-fixated specimens were provided through the local anatomy department. We defined moderate-to-severe cartilage degeneration, such as substantial tissue loss or focal lesions, as an exclusion criterion. Before MR image acquisition, the specimens were left to thaw at room temperature for 24 hours.

## MR Image Acquisition

The MIXTURE technique uses interleaved 3D TSE sequence blocks of two or more image weightings, resulting in two (or more) differently weighted stacks of 3D TSE images. Consequently, quantitative maps are reconstructed for each slice. Image weightings are freely adjustable by selecting a preparation module, i.e., none, T2 preparation, or T1ρ preparation, and prepulse, i.e., none, inversion recovery (IR), or spectral attenuated inversion recovery (SPAIR) (**Figure 1a**). Inspired by our clinical protocol of 2D TSE sequences, i.e., PD-w FS and T1-w sequences, two MIXTURE sequences were set up: (i) MIXTURE PD-w FS with T2 mapping (termed 'MIX1') and (ii) MIXTURE T1-w with T1ρ mapping (termed 'MIX2').

For MIX1, as shown in **Figure 1b**, two TSE blocks were interleaved using a repetition time (TR) of 1200 ms. The first TSE block was configured without any relaxation preparation module, i.e., setting the echo time (TE) of the T2-preparation module to 0 ms, but with a SPAIR prepulse, resulting in PD-w FS images. The second TSE block was configured without a SPAIR prepulse but with a T2 preparation module using a TE of 50 ms, resulting in T2-w images. Based on the two images, T2 maps were reconstructed on the scanner workstation by fitting a mono-exponential signal decay function to the two TE values in a voxel-wise manner.

For MIX2, as shown in **Figure 1c**, three TSE blocks were interleaved using a TR of 600 ms. The first TSE block was configured without a SPAIR prepulse and by setting the spin lock time (TSL) of the T1ρ-preparation module to 0 ms, resulting in T1-w images. The second and third blocks were configured with a SPAIR prepulse and T1ρ preparation modules of TSL = 25 ms and TSL = 50 ms, respectively, at an offset frequency of 500 Hz, resulting in T1ρ-w FS images. Based on the three images, T1ρ maps were reconstructed as detailed above.

Our standard protocol's PD-w FS and T1-w 2D TSE sequences and the corresponding 3D TSE sequences were acquired using the same basic image acquisition parameters, i.e., slice orientation, slice thickness, number of slices, acquisition matrix, and reconstruction matrix, to permit voxel-to-voxel comparisons. Consequently, despite their 3D nature, the MIXTURE and the 3D TSE sequences were acquired non-isotropically with lower through-plane resolution (i.e., thicker slices) and higher in-plane resolution than previously published variants [14] [15] [16].

**Table 1** summarizes the sequence parameters.

## Qualitative and Semi-quantitative Analysis

Three clinical radiologists (███, ███, ██ BLINDED FOR REVIEW ONLY) with 3, 4, and 4 years of clinical experience rated the images on in-house radiology workstations. The images were presented head-to-head using the in-house picture archiving and communication (PACS) system (iSite, Philips, The Netherlands). MIX 1 (i.e., the MIXTURE PD-w FS image, MIXTURE T2-w image [TE=50 ms], and MIXTURE T2 maps) were presented alongside the corresponding 2D and 3D TSE PD-w FS images. Analogously, MIX 2 (i.e., the MIXTURE T1-w image, MIXTURE T1ρ-w FS image [TSL=50 ms], and MIXTURE T1ρ maps) were presented alongside the corresponding 2D and 3D TSE T1-w images. The parameter maps had been prepared as video files in Python (version 3.9.9), using the “jet” color map and fixed ranges (0ms - 100ms [T2 map], 0ms - 70ms [T1ρ map]), to allow for scrolling on a separate screen.

The radiologists semiquantitively rated the images on a per-sequence and per-joint basis. The diagnostic evaluability of different knee joint structures was quantified using an ordinal 5-point Likert scale, where score 1 indicated ‘very low evaluability’, score 2 ‘low evaluability’, score 3 ‘intermediate evaluability’, score 4 ‘high evaluability’, and score 5 ‘very high evaluability’. These structures were selected as suggested by Chaudhari et al. [13]. The radiologists assigned lower scores to structures they would usually not primarily evaluate on a given sequence or orientation. In addition, image quality, contrast resolution, and anatomic delineation were synoptically assessed as a global diagnostic quality score for each sequence on a per-joint basis, using a similar ordinal 5-point Likert scale that extended from ‘very poor’ (score 1) to ‘very good’ (score 5). Also, the radiologists were instructed to note the presence of artifacts (i.e., suspicious features or distortions) in the morphologic 3D TSE or MIXTURE images compared to the 2D TSE images. Consequently, blinding to sequence type was not feasible. Besides, the radiologists indicated whether the availability of the quantitative parameter map had increased their diagnostic confidence.

## Quantitative Analysis

Region of interest (ROI)-based comparisons between sequences were similarly performed on a per-joint and per-sequence basis. ROI placement was standardized for all sequences: █, a pre-graduate medical student with two years of experience in MRI, placed the ROIs, which █, a medical imaging scientist with eight years of experience in MRI, reviewed subsequently.

### Image Contrast

To evaluate image contrast and contrast-to-noise ratio (CNR), two circular ROIs with a standard area (2.11 $mm^2$) were positioned in predefined anatomic structures. In the PD-w FS images, ROIs were placed within the femoral cartilage (central weight-bearing region of the lateral compartment) and the synovial fluid of the patellofemoral and femorotibial joint compartment. In the T1-w images, ROIs were placed within the femoral cartilage and the central infrapatellar fat pad. Weber contrast [17] was used as a surrogate of image contrast and calculated based on the signal intensity of cartilage versus synovial fluid or fatty tissue:

$$C_w{}^{PDw\,FS} = \frac{SI_{cart} - SI_{syn}}{SI_{syn}}$$

and

$$C_w^{T_1w} = \frac{SI_{cart} - SI_{fat}}{SI_{fat}},$$

where $SI_{cart}$, $SI_{syn}$, and $SI_{fat}$ are the mean signal intensities of the cartilage, synovial fluid, and fat pad ROIs.

CNR was used to quantify how distinguishable two structures are from the noise floor. Higher CNR values indicate that the two structures are more clearly distinguishable. CNR was calculated as

$$CNR^{PDw\,FS} = \frac{|SI_{cart} - SI_{syn}|}{\sqrt{\sigma_{cart}^2 + \sigma_{syn}^2}}$$

and

$$CNR^{T_1w} = \frac{|SI_{cart} - SI_{fat}|}{\sqrt{\sigma_{cart}^2 + \sigma_{fat}^2}},$$

respectively, where $\sigma_{cart}$, $\sigma_{syn}$, and $\sigma_{fat}$ are the standard deviations of the cartilage, synovial fluid, and fat pad ROIs.

## Signal Quality

As a surrogate for the signal-to-noise ratio, the coefficient of variation (CV) was quantified [18]. A circular ROI with an area of 3.76 mm² was positioned in a homogeneous area of the synovial fluid (PD-w FS images) or infrapatellar fat pad (T1-w images). CV was calculated as

$$CV^{PD-w\,FS} = \frac{\sigma_{syn}}{SI_{syn}}$$

and

$$CV^{T1-w} = \frac{\sigma_{fat}}{SI_{fat}},$$

respectively. Higher CV values indicate stronger noise. No CV was calculated for cartilage because of its distinct layer structure that is likely to make tissue inhomogeneity dominate over the statistical signal fluctuations.

### Cartilage Segmentation

The femoral and tibial cartilage plates were segmented on the central slice of the lateral compartment based on the sagittal MIXTURE PD-w FS images. The segmentations were performed by ██ using the Program ITK-Snap (v3.6, [19]), and reviewed by ██, a musculoskeletal radiologist with more than ten years of experience in MSK radiology.

## Statistical Analysis

NP, TL, KLR, and SN performed the statistical analyses. Separate inter-sequence comparisons between MIXTURE, 2D TSE, and 3D TSE were done for the PD-w FS and the T1-w images. Diagnostic evaluability scores and the global diagnostic quality scores were analyzed in R (v4.0.3, R Foundation for Statistical Computing) using mixed-effects ordinal regression, i.e., the Cumulative Link Mixed Model (CLMM) from the 'ordinal' package. The sequence was the independent variable, while structure, reader, and joint specimen were set as random variables. The model was fitted using the Laplace approximation. Another CLMM was fitted to compare the global diagnostic quality scores between sequences. For both models, estimated marginal means (EMM) were calculated and compared post hoc using TukeyÄs multiplicity adjusted post-hoc test. EMMs represent the model-predicted response for each level of a variable, while averaging over other variables [20]. In the context of CLMMs, EMM values are defined on a latent scale [21], i.e., they do not directly relate to the ordinal response variable. Still, higher EMM values indicate higher values of the respective variable. Weber contrast, CNR, and CV were compared using Graph Pad Prism software (v9.5.1, San Diego, CA, USA) and repeated measures analysis of variance (ANOVA) followed by the Tukey-Kramer post-hoc test. Multiplicity-adjusted p-values were computed to account for multiple comparisons. A family-wise alpha error threshold of $p \leq 0.01$ was chosen to limit the number of statistically significant, yet clinically (most likely) non-significant findings.

## Results

*Study cohort*

All ten knee joint specimens (age 81.1±10.4 years [mean±standard deviation]; range 68–96 years; 9/1 male/female; 6/4 right/left joints) could be imaged per the study protocol.

*PD-weighted FS images*

Across the acquired sequences, cartilage demonstrated an intermediate signal intensity and a layered zonal structure (**Figures 2 and 3**). However, the ROI-based analysis yielded a significantly more pronounced contrast as evidence by significantly more negative Weber contrast values between cartilage and synovia for the 2D TSE images ($p<0.001$ [**Figure 4a**]). In visual terms, the negative Weber contrast values translate into a darker presentation of the cartilage tissue than the bright synovial fluid. Similarly, CNR values tended to be higher in the 2D TSE images, though not significantly (**Figure 4b)**, and CV values were largely similar (**Figure 4c**). Quantitative T2 maps of the femoral and tibial cartilage indicated a zonal stratification of the T2 relaxation times with decreasing values along increasing tissue depth (**Figure 3e, f)**.

*T1-weighted images*

Across the acquired sequences, contrast, signal, and microstructural details appeared largely similar, although slightly more blurring was observed for the MIXTURE T1-w images (**Figures 5 and 6**).

Consequently, Weber contrast, CNR, and CV values were not significantly different (**Figure 4d-f**). Quantitative T1ρ maps of the cartilage indicated a depth-related zonal stratification, too (**Figure 6e,f**).

*Diagnostic evaluability and global diagnostic quality scores – PD-weighted FS images*

Except for the collateral ligaments, the radiologists assigned evaluability scores of 4 or 5 on PD-w FS sequences, indicating 'high' to 'very high' evaluability with only slight inter-sequence differences (**Table 2**). Global diagnostic quality was largely rated as 'very good' with median values of 5 [5; 5] (median [lower quartile; upper quartile]) for MIXTURE and 2D TSE PD-w FS vs. 5 [4; 5] for 3D TSE PD-w FS. One radiologist felt the additional T2 maps increased their diagnostic confidence in seven of ten assessed knee joints, while the other two radiologists considered them valuable in one and zero knee joints only, respectively. For the diagnostic evaluability scores, the highest EMM values, indicating overall higher scores, were found for the MIXTURE sequence ($5.4 \pm 0.8$ [EMM±standard error]), compared to the corresponding 2D TSE ($5.1 \pm 0.8$) and 3D TSE ($4.7 \pm 0.8$) sequences. Pair-wise post-hoc tests indicated significantly higher diagnostic evaluability scores for MIXTURE than 3D TSE ($P<0.001$). In contrast, no significant differences were found between 2D TSE and MIXTURE ($P=0.055$) and 2D TSE and 3D TSE ($P=0.019$). Evaluation of the global diagnostic quality score yielded EMM values of $7 \pm 3$, $10 \pm 3$, and $4 \pm 3$ for the MIXTURE, 2D TSE, and 3D TSE sequences, respectively, which were not significantly different (**Table 3**).

*Diagnostic evaluability and global diagnostic quality scores – T1-weighted images*

The radiologists assigned an evaluability score of 5 to the bone marrow, while other structures mainly scored 3, indicating 'intermediate evaluability', while the collateral ligaments were scored lower (**Table 2**). Nonetheless, the global diagnostic quality of all sequences was rated with a median score of 5, while only subtle differences were found for the lower quartiles. The radiologists' judgment on the potential of the MIXTURE T1ρ maps to increase diagnostic confidence was ambiguous as they reported additional diagnostic value in one, three, or nine assessed knee joints. The highest EMM values were found for the MIXTURE sequence (1.1±0.6), which were significantly higher than those for the 2D TSE (0.3±0.6) and 3D TSE (0.3±0.6) sequences ($p<0.001$). The EMM values of the 2D TSE and 3D TSE sequences did not differ significantly ($p=0.999$). Evaluation of the global diagnostic quality score yielded EMM values of 2±2, 3±2, and 1±2 for the MIXTURE, 2D TSE, and 3D TSE sequences, respectively, which did not differ significantly (**Table 3**).

*Artificial findings*

In two knee joint specimens, artificial findings of the meniscal posterior horns were noted for the MIXTURE and 3D TSE sequences, i.e., a fine hyperintense line extending from the meniscus base to the red-white zone that may principally be mistaken as a tear and a hyperintense area toward the meniscal undersurface that could be mistaken as degenerative meniscopathy. We consider these signal alterations artificial because they were absent in the corresponding 2D TSE and MIXTURE T2-weighted sequences (**Figure 7**). Yet, correlates were visible on the respective T1-w images.

# Discussion

This study developed and refined two MIXTURE sequence variants combining PD-w FS or T1-w morphologic images with T2 mapping or T1ρ mapping, respectively. The variants matched our institution's knee protocol regarding image weightings and provided T2 and T1ρ maps of the whole joint in clinically feasible scan times. The most important finding of our study is that MIXTURE sequences were assigned similar or even significantly higher diagnostic evaluability and quality scores – both globally and per structure.

The MIXTURE platform is versatile because different image weightings can be included by selecting from different preparation modules. Other weightings beyond the ones chosen in this study are feasible to reflect different practices. Previously, Sakai et al. combined PD-w images without fat saturation with T2-w FS images and T2 maps [15]. However, a MIXTURE sequence with FS yields meaningful T2 or T1ρ relaxation times only in non-fatty structures. Intentionally, we built non-isotropic MIXTURE variants, allowing for voxel-wise comparisons between the 2D and 3D sequences. However, while scientifically well-justified, this approach also meant that we could not perform slice reconstruction in arbitrary orientations.

The proposed MIXTURE, 2D TSE, and 3D TSE sequences were similar in contrast-to-noise ratio, contrast, and coefficient of variation. While remaining contrast differences are likely not diagnostically detrimental because signal intensities can be manually adjusted, they result most from different sequence parameters. For example, the MIXTURE PD-w FS sequence used a shorter TR than the 2D TSE sequence to minimize the scan time yet was combined with a specific refocusing pattern to realize the familiar image appearance.

We used human cadaveric knee joint specimens for our evaluation. Compared to actual patient measurements, this approach limits the clinical translation of our findings. However, we found this model optimal for inter-sequence comparisons as the potential inter-sequence motion was effectively excluded, thereby improving standardization and comparability. Our radiologists noticed that the T2-weighted MIXTURE images looked “unnaturally” dark. This finding is plausible as human tissues, when imaged at room temperature, have substantially reduced contrast between fat and muscle (on T2-w images), while the contrast between fat and fluid increases [22]. When acquired in a healthy volunteer, the images demonstrated familiar characteristics and contrasts.

Quantitatively, we found similar (per-structure) evaluability and (global) diagnostic quality scores between the sequences. Qualitatively, the readers’ opinions on the utility of the quantitative parameter maps were ambiguous. While one radiologist considered them largely beneficial, the other two reported the opposite, which may reflect that (i) clinical diagnosis still relies exclusively on morphologic sequences and (ii) new sequences need familiarization (beyond mere clinical adoption) to be considered beneficial. To fit into clinical workflows, the quantitative parameter maps must be amenable to automated segmentation approaches, which could be facilitated by isotropic morphologic MIXTURE images. Computer-aided and deep-learning enhanced segmentation models may be available in the near future [23] [24].

In the MIXTURE and 3D TSE PD-w FS sequences, we observed hyperintense intra-meniscal signal alterations in two knee joints. These signal alterations were absent in the 2D TSE sequence, indicating their artificiality. While MIXTURE T2-w images confirmed these as artificial, their cause remains uncertain. We speculate that residual T1 contrast may have produced these artifacts, evidenced by hyperintense correlates on the T1-w images and possibly amplified by the 3D sequence refocusing patterns and shorter TRs (of the MIXTURE and 3D TSE).

Our study has limitations. First, the specimen size and type were limited as cadavers are not fully representative of the conditions in patients. Similarly, pathological variability was limited, and consequently, we consider this study a proof-of-concept that needs to be followed by future patient studies. Second, the design of our reader study ignored that some structures within the knee joint are routinely judged based on PD-w FS or T1-w sequences only. Radiologists were instructed to subjectively assess a given structure and assign lower scores if a structure was poorly evaluable. Most intra-articular structures, such as the menisci and cartilage, are routinely evaluated on the PD-w FS sequences, which explains their higher scores. Third, we only evaluated one clinical 3T MRI scanner and coil. The sequence's usability for different MRI scanners, field strengths, and coils remains to be studied. Fourth, we could not validate MIXTURE's T2 and T1ρ relaxation times against nominal relaxation times obtained by reference sequences. Ideally, a phantom with known relaxation times [25] and appropriate geometry (to fit into the knee coil) would be used to study the validity of the quantitative approaches versus reference sequences such as multi-echo-spin echo T2 mapping sequences and alternative combined sequences such as qDESS. Yet, most commercially available phantoms are head-sized. Also, there is no consensus about T1ρ reference sequences yet, because of multiple technical confounders that may affect both MIXTURE and reference sequences [26]. Fifth, this study could not evaluate the susceptibility of MIXTURE sequences to blood flow and motion because of the study's in-situ nature.

In conclusion, the MIXTURE sequence platform provides (i) morphologic images of diagnostic quality and adjustable TSE-based contrasts to be aligned with a radiologist’s or institution’s practices and (ii) quantitative parameter mapping with whole-joint coverage and additional insights on soft tissue composition and ultrastructure in clinically feasible scan times. By increasing diagnostic efficiency, MIXTURE provides a versatile approach to include quantitative MRI techniques in clinical routine protocols.

# References

[1] OARSI Osteoarthritis Research Society International, „Osteoarthritis: A Serious Disease, Submitted to the U.S. Food and Drug Administration December 1, 2016“.

[2] H. J. Braun und G. E. Gold, „Diagnosis of osteoarthritis: Imaging,“ *Bone,* Nr. 51, p. 278–288, 2012.

[3] C. Y. J. Wenham, A. J. Grainger und P. G. Conaghan, „The role of imaging modalities in the diagnosis, differential diagnosis and clinical assessment of peripheral joint osteoarthritis,“ *Osteoarthritis and Cartilage,* Bd. 22, Nr. 10, pp. 1692-1702, 2014.

[4] A. Guermazi, H. Alizai, M. Crema, S. Trattnig, R. Regatte und F. Roemer, „Compositional MRI techniques for evaluation of cartilage degeneration in osteoarthritis,“ *Osteoarthritis and Cartilage,* Bd. 23, Nr. 10, pp. 1639-1653, 2015.

[5] D. Binks, R. Hodgson, M. Ries, R. Foster, S. Smye, D. McGonagle und A. Radjenovic, „Quantitative parametric MRI of articular cartilage: a review of progress and open challenges,“ *British Journal of Radiology,* Bd. 86, p. 20120163, 2013.

[6] J. Thüring, K. Linka, M. Itskov, M. Knobe, L. Hitpaß, C. Kuhl, D. Truhn und S. Nebelung, „Multiparametric MRI and Computational Modelling in the Assessment of Human Articular Cartilage Properties: A Comprehensive Approach,“ *BioMed Research International,* p. 9460456, 2018.

[7] J. Le, Q. Peng und K. Sperling, „Biochemical magnetic resonance imaging of knee articular cartilage: T1rho and T2 mapping as cartilage degeneration biomarkers,“ *Annals of the New York Academy of Sciences,* Bd. 1383, Nr. 1, pp. 34-42, 2016.

[8] B. Hager, M. Raudner, V. Juras, O. Zaric, P. Szomolanyi, M. Schreiner und S. Trattnig, „MRI of Early OA,“ in *Early Osteoarthritis: State-of-the-Art Approaches to Diagnosis, Treatment and Controversies*, Cham, Springer, 2022, pp. 17-26.

[9] R. Kijowski, D. G. Blankenbaker, d. M. A. Rio, G. S. Baer and B. K. Graf, "Evaluation of the Articular Cartilage of the Knee Joint: Value of Adding a T2 Mapping Sequence to a Routine MR Imaging Protocol," *Radiology,* no. 267(2), pp. 503-513, 2013.

[10] M. A. I. Alsayyad, K. A. A. Shehata und R. T. Khattab, „Role of adding T2 mapping sequence to the routine MR imaging protocol in the assessment of articular knee cartilage in osteoarthritis,“ *Egyptian Journal of Radiology and Nuclear Medicine,* Bd. 52, Nr. 78, 2021.

[11] M. C. Nevitt, D. T. Felson and G. Lester, "The Osteoarthritis Initiative: Protocol for the Cohort Study," [Online]. Available: https://nda.nih.gov/static/docs/StudyDesignProtocolAndAppendices.pdf. [Accessed 2 September 2022].

[12] A. Chaudhari, K. Stevens, B. Sveinsson, J. Wood, C. Beaulieu, E. Oei, J. Rosenberg, F. Kogan, M. Alley, G. Gold und B. Hargreaves, „Combined 5-minute double-echo in steady-state with separated echoes and 2-minute proton-density-weighted 2D FSE sequence for comprehensive

whole-joint knee MRI assessment.," *J Magnetic Resonance Imaging,* Bd. 49, Nr. 7, pp. 183-194, 2019.

[13] A. S. Chaudhari, M. J. Grissom, Z. Fang, B. Sveinsson, J. H. Lee, G. E. Gold, B. A. Hargreaves und K. J. Stevens, „Diagnostic Accuracy of Quantitative Multicontrast 5-Minute Knee MRI Using Prospective Artificial Intelligence Image Quality Enhancement," *American Journal of Roentgenology,* Nr. 216, pp. 1-12, 2021.

[14] M. Yoneyama, T. Sakai, S. Zhang, D. Murayama, H. Yokota, Y. Zhao, S. Saruya, M. Suzuki, A. Watanabe, M. Niitsu und M. Van Cauteren, „MIXTURE: A novel sequence for simultaneous morphological and quantitative imaging based on multi-interleaved 3D turbo-spin echo MRI," in *Proc. ISMRM 2021:4203*, 2021.

[15] T. Sakai, M. Yoneyama, A. Watanabe, D. Murayama, S. Ochi, S. Zhang und T. Miyati, „Simultaneous anatomical, pathological and T2 quantitative knee imaging with 3D submillimeter isotropic resolution using MIXTURE," in *Proc. ISMRM 2021: 0845*, 2021.

[16] S. Saruya, M. Suzuki, M. Yoneyama, K. Inoue, E. Kozawa und M. Niitsu, „3D sub-millimeter isotropic knee cartilage T1rho mapping using multi-interleaved fluid-attenuated TSE acquisition (MIXTURE)," in *Proc. ISMRM 2021: 2977*, 2021.

[17] S. Gupta und R. Porwal, „Appropriate Contrast Enhancement Measures for Brain and Breast Cancer Images," *Int J Biomed Imaging,* p. 4710842, 2016.

[18] J. Weiss, M. Notohamiprodjo, P. Martirosian, J. Taron, M. D. Nickel, M. Kolb, F. Bamberg, K. Nikolaou und A. E. Othman, „Self-Gated 4D-MRI of the Liver: Initial Clinical Results of Continuous Multiphase Imaging of Hepatic Enhancement," *Journal of Magnetic Resonance Imaging,* Bd. 47, pp. 459-467, 2018.

[19] P. A. Yushkevich, J. Piven, H. C. Hazlett, R. G. Smith, S. Ho, J. C. Gee und G. Gerig, „User-guided 3D active contour segmentation of anatomical structures: Significantly improved efficiency and reliability," *Neuroimage,* Bd. 31, Nr. 3, pp. 1116-28, 2006.

[20] [Online]. Available: http://cran.nexr.com/web/packages/emmeans/vignettes/basics.html. [Accessed 27 09 2023].

[21] [Online]. Available: https://cran.r-project.org/web/packages/emmeans/vignettes/sophisticated.html#ordinal.

[22] T. Ruder, M. Thali und G. Hatch, „Essentials of forensic post-mortem MR imaging in adults," *British Journal of Radiology,* Bd. 87, p. 20130567, 2014.

[23] J. Schock, M. Kopaczka, B. Agthe, J. Huang, P. Kruse, D. Truhn, S. Conrad, G. Antoch, C. Kuhl, S. Nebelung und D. Merhof, „A Method for Semantic Knee Bone and Cartilage Segmentation with Deep 3D Shape Fitting Using Data from the Osteoarthritis Initiative," in *Shape in Medical Imaging. ShapeMI 2020. Lecture Notes in Computer Science()*, Bd. 12474, M. Reuter, C. Wachinger, H. Lombaert, B. Paniagua, O. Goksel und I. Rekik, Hrsg., Cham, Springer, 2020.

[24] A. D. a. B. M. a. M. V. a. R. E. a. B. M. S. a. W. L. E. a. G. G. E. a. H. B. A. a. C. A. S. Desai, „DOSMA: A deep-learning, open-source framework for musculoskeletal MRI analysis," in *Proc 27th Annual Meeting ISMRM*, Montreal, 2019.

[25] N. A. Obuchowski, A. P. Reeves, E. P. Huang, X.-F. Wang, A. J. Buckler, H. J. (. Kim, H. X. Barnhart, E. F. Jackson, M. L. Giger, G. Pennello, A. Y. Toledano, J. Kalpathy-Cramer, T. V. Apanasovich und e. al., „Quantitative imaging biomarkers: A review of statistical methods for computer algorithm comparisons," *Statistical Methods in Medical Research,* Bd. 24, Nr. 1, p. 68–106, 2015.

[26] W. Chen, „Errors in quantitative T1rho imaging and the correction methods," *Quantitative Imaging in Medicine and Surgery,* Bd. 5, Nr. 4, pp. 583-591, 2015.

# Figures

## Figure 1

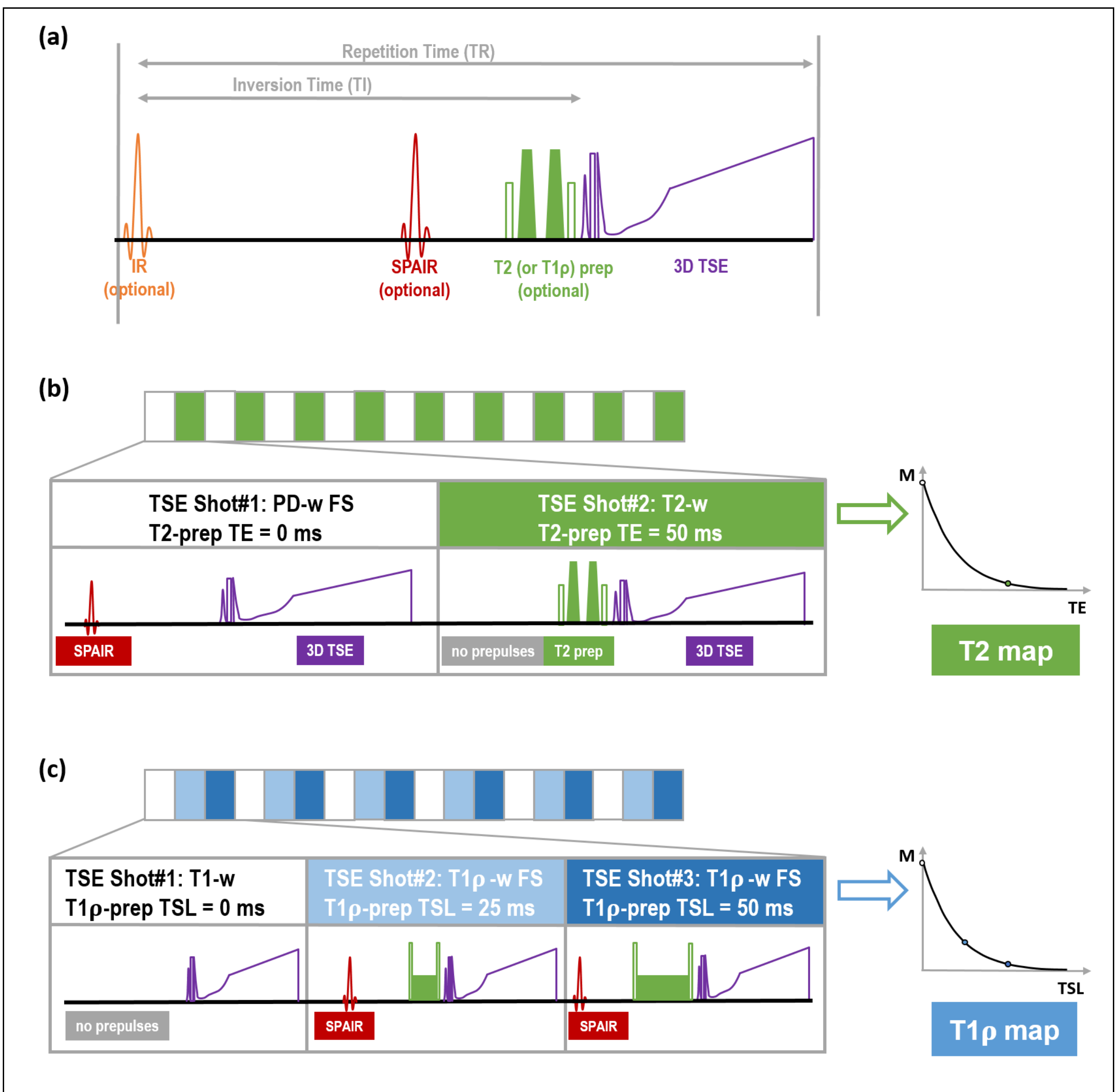

**Figure 1: MIXTURE sequence platform.** The MIXTURE sequence platform allows 3D TSE imaging with flexible preparation.
(a) The basic building block of the sequence consists of optional T2 (or T1ρ) preparation (prep) modules and optional inversion recovery and/or SPAIR fat saturation prepulse to realize variable contrast weightings.
(b) The MIXTURE PD-weighted (-w) fat-saturated (FS) sequence with $T_2$ mapping interleaves TSE shots without (white blocks) and with T2 preparation (green blocks). In the PD-w FS block, a SPAIR prepulse is activated. Based on the resulting PD-w FS and T2-w images, T2 maps are calculated.
(c) The MIXTURE T1-w sequence with T1ρ mapping interleaves TSE shots without (white blocks) and with T1ρ preparation modules of different spin lock times (TSL) (light blue and dark blue blocks). In the T1ρ-w FS blocks, a SPAIR prepulse is activated. Based on the resulting T1-w and T1ρ-w FS images, a T1ρ map is calculated.

Abbreviations: MIXTURE - Multi-Interleaved X-prepared Turbo-Spin Echo with IntUitive Relaxometry, SPAIR - Spectral Attenuated Inversion Recovery.

## Figure 2

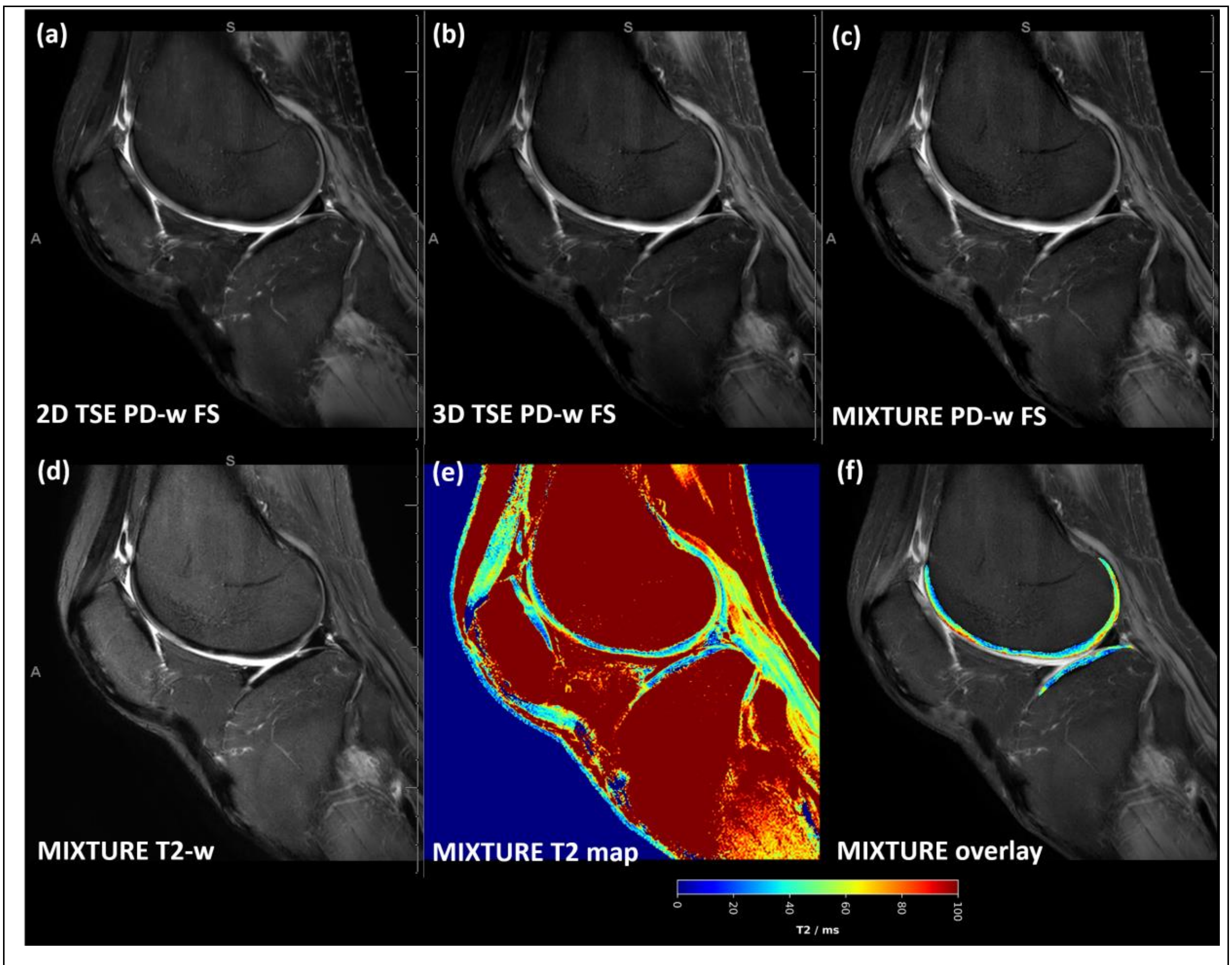

**Figure 2: Proton density-weighted fat-saturated images acquired using the MIXTURE and corresponding reference sequences.**

For a representative slice (through the central lateral femorotibial compartment) and joint, the 2D TSE image (a), the 3D TSE image (b), and the MIXTURE image (c) are shown alongside the MIXTURE T2-weighted image (d), the MIXTURE T2 map (e), and the segmented cartilage tissue overlay (f). In this example, the segmented femoral and tibial cartilage exhibited T2 relaxation times of 48±18 ms and 36±16 ms, respectively (mean ± standard deviation). In areas of fatty tissue, the MIXTURE T2 map did not yield meaningful values because SPAIR fat saturation was employed during the acquisition of the PD-weighted FS morphologic images. Abbreviations: PD – proton density, -w – weighted, FS – fat-saturated, MIXTURE - Multi-Interleaved X-prepared Turbo-Spin Echo with IntUitive Relaxometry, SPAIR - Spectral Attenuated Inversion Recovery. **Figure 3** provides a close-up of the femoral and tibial cartilage of the weight-bearing region.

## Figure 3

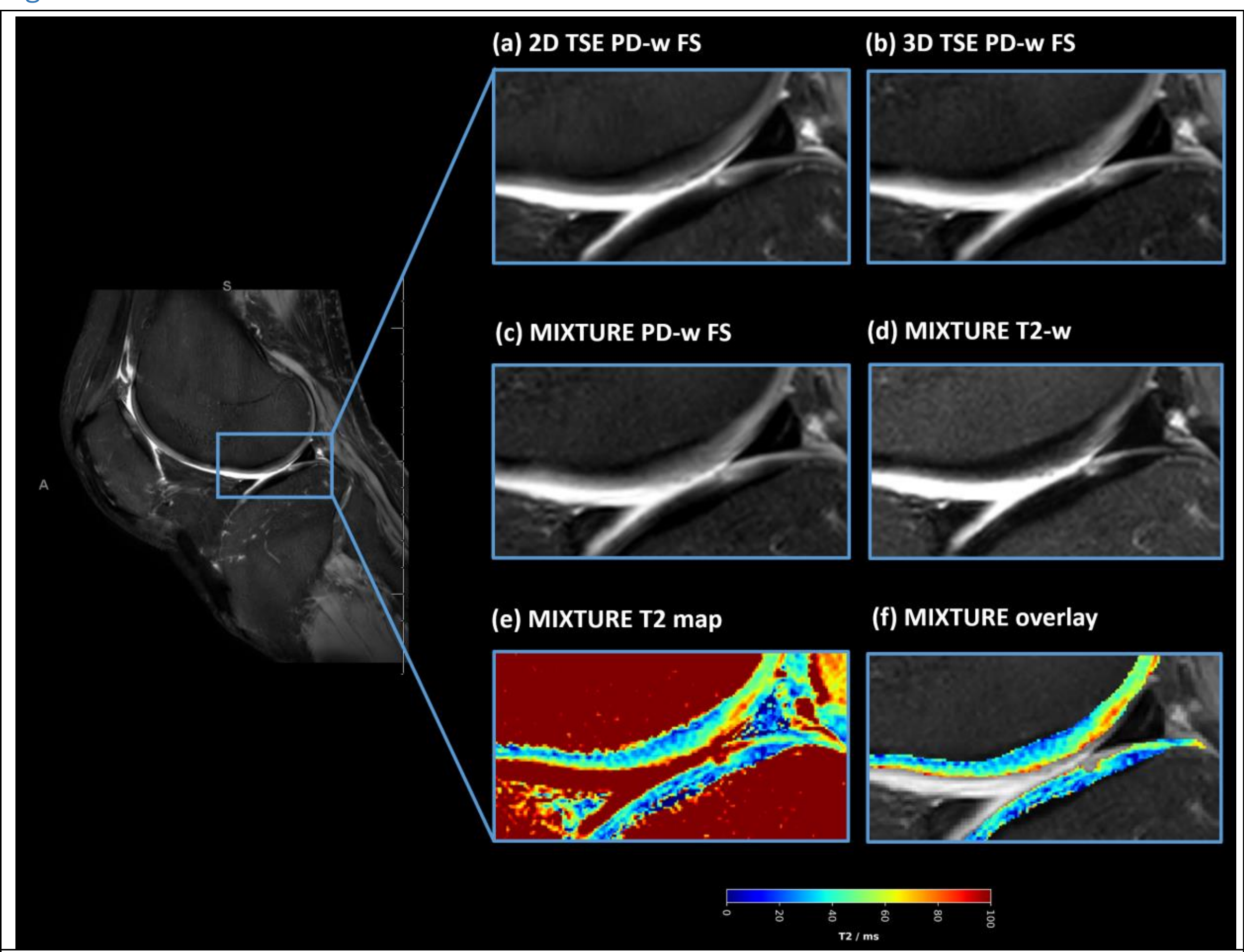

**Figure 3: Close-up of the central weight-bearing joint region.**
The blue box indicates the zoomed-in area. Figure organization, specimen, and slice as in **Figure 2**.

Figure 4

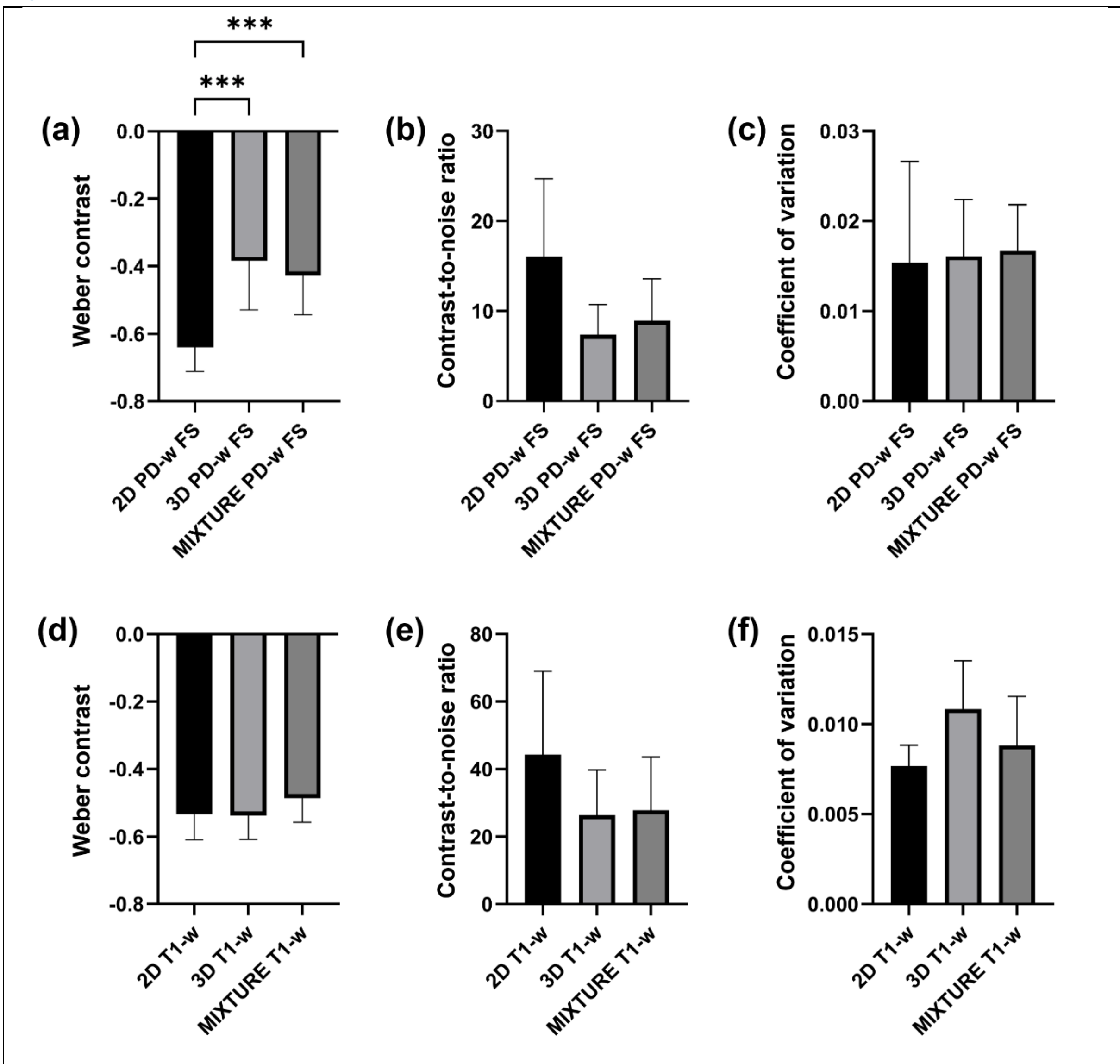

**Figure 4: Metrics of image contrast and signal quality as a function of sequence and weighting.** ROI-based comparisons were performed for the proton density-weighted (PD-w) fat-saturated (FS) sequences ((a) to (c)) and the T1-w sequences ((d) to (f)).

(a,d) Weber contrast between cartilage and synovia (a) and cartilage and the infrapatellar fat pad (d).

(b,e) Contrast-to-noise ratios between cartilage and synovia (b) and cartilage and the infrapatellar fat pad (e).

(c,f) Coefficients of variation in synovial fluid (c) and the infrapatellar fat pad (f).

Asterisks "**" and "***" indicate multiplicity-adjusted p-values of $0.001<p\leq0.01$ and $p\leq0.001$, respectively. Bars and whiskers indicate means and standard deviations.

## Figure 5

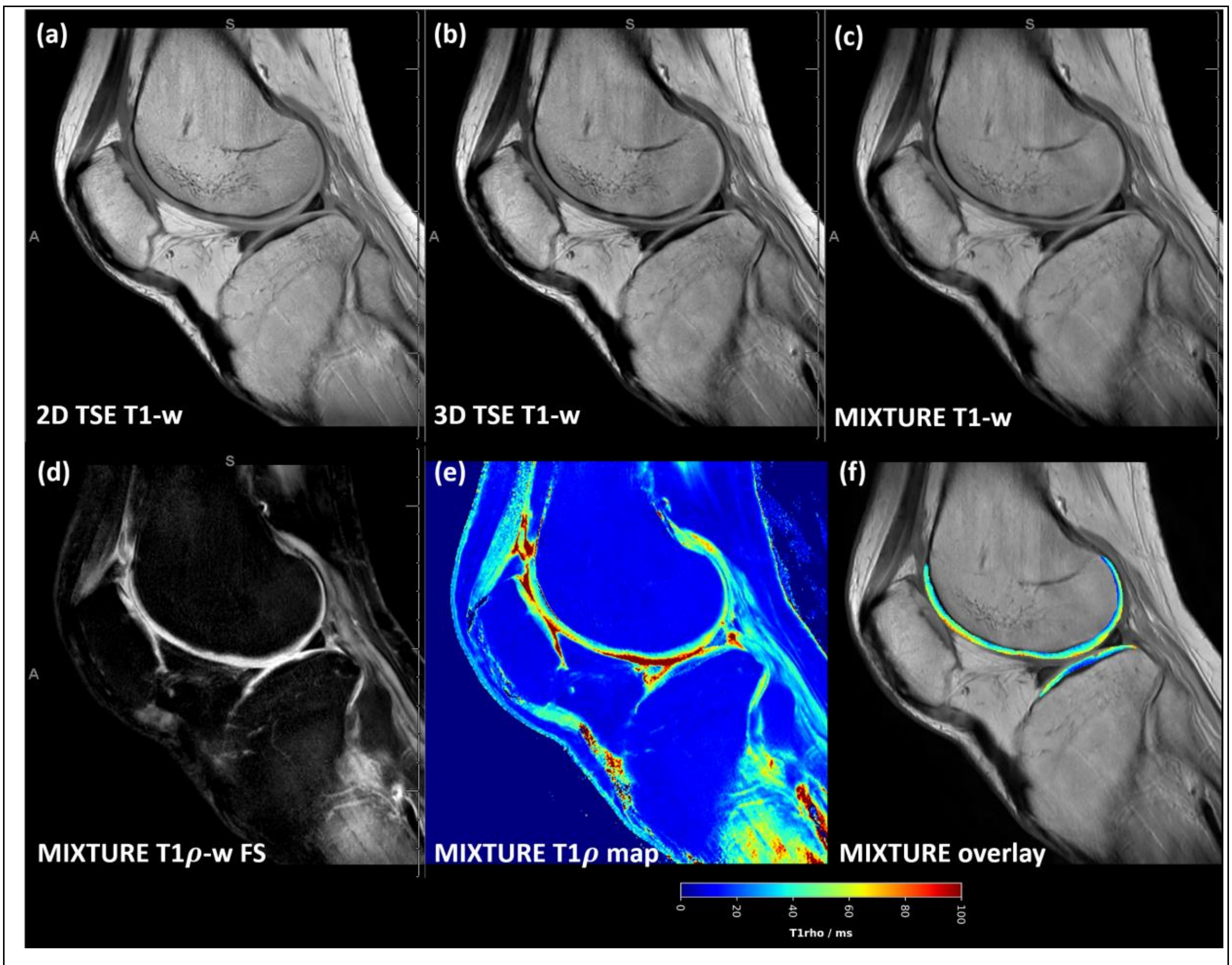

**Figure 5: T1-weighted images acquired using the MIXTURE and corresponding reference sequences.**

The 2D TSE image (a), the 3D TSE image (b), and the MIXTURE image (c) are shown alongside the MIXTURE T1ρ-weighted fat-saturated image with 50 ms spin lock time (d), the MIXTURE T1ρ map (e), and the segmented cartilage tissue overlay (f). In this example, the segmented area of femoral and tibial cartilage exhibited mean T1ρ relaxation times of 45±14 ms and 38±14 ms, respectively (mean ± standard deviation). As for the MIXTURE T2 maps (**Figures 2 and 3**), the MIXTURE T1ρ map did not yield meaningful values in areas of fatty tissue. Same slice and joint as in **Figure 2**. Abbreviations: -w – weighted, FS – fat saturated. MIXTURE - Multi-Interleaved X-prepared Turbo-Spin Echo with IntUitive Relaxometry. **Figure 6** provides a close-up of the femoral and tibial cartilage of the weight-bearing region.

## Figure 6

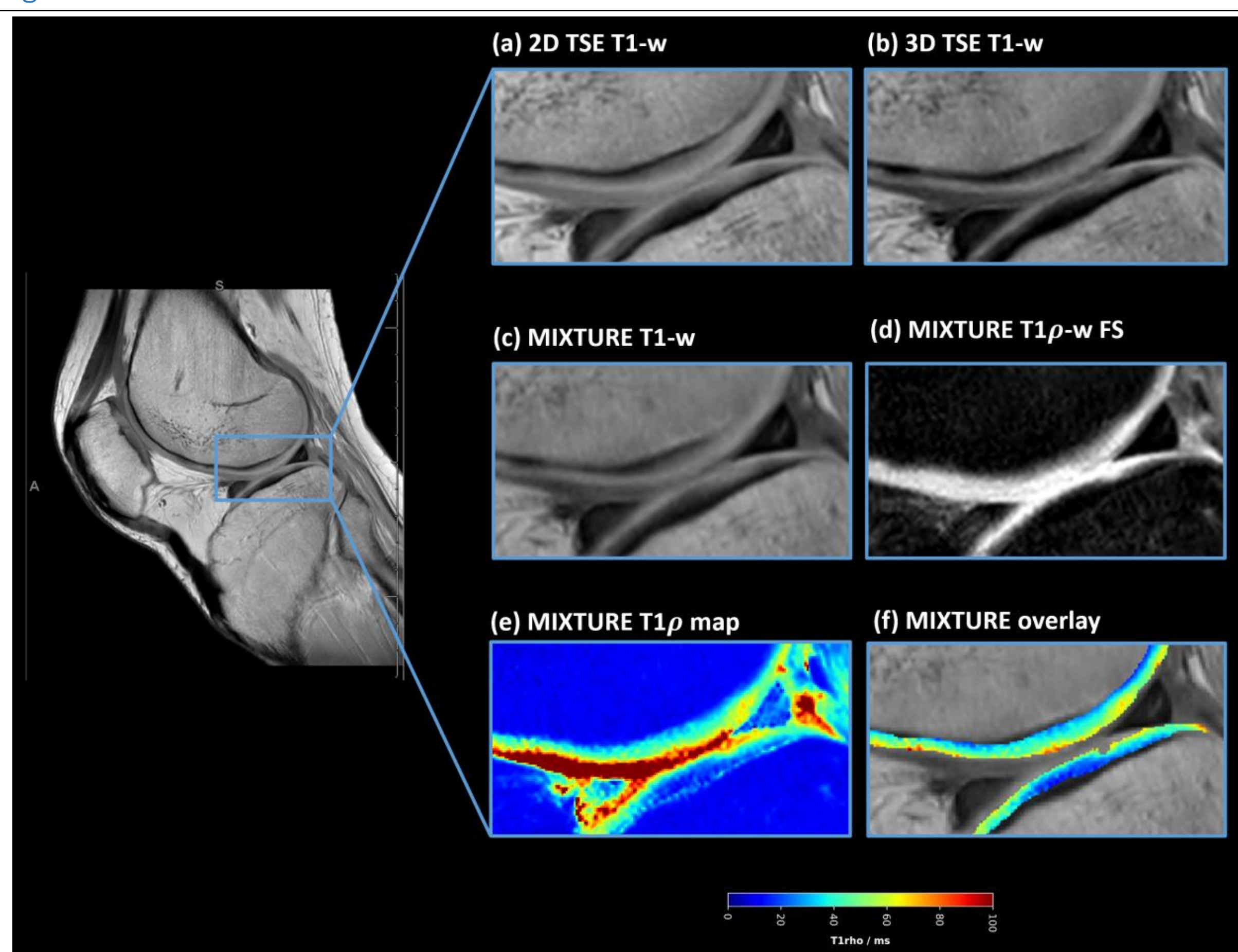

**Figure 6: Close-up of the central weight-bearing joint region.**
The blue box indicates the zoomed-in area. Figure organization, specimen, and slice as in **Figure 3**.

## Figure 7

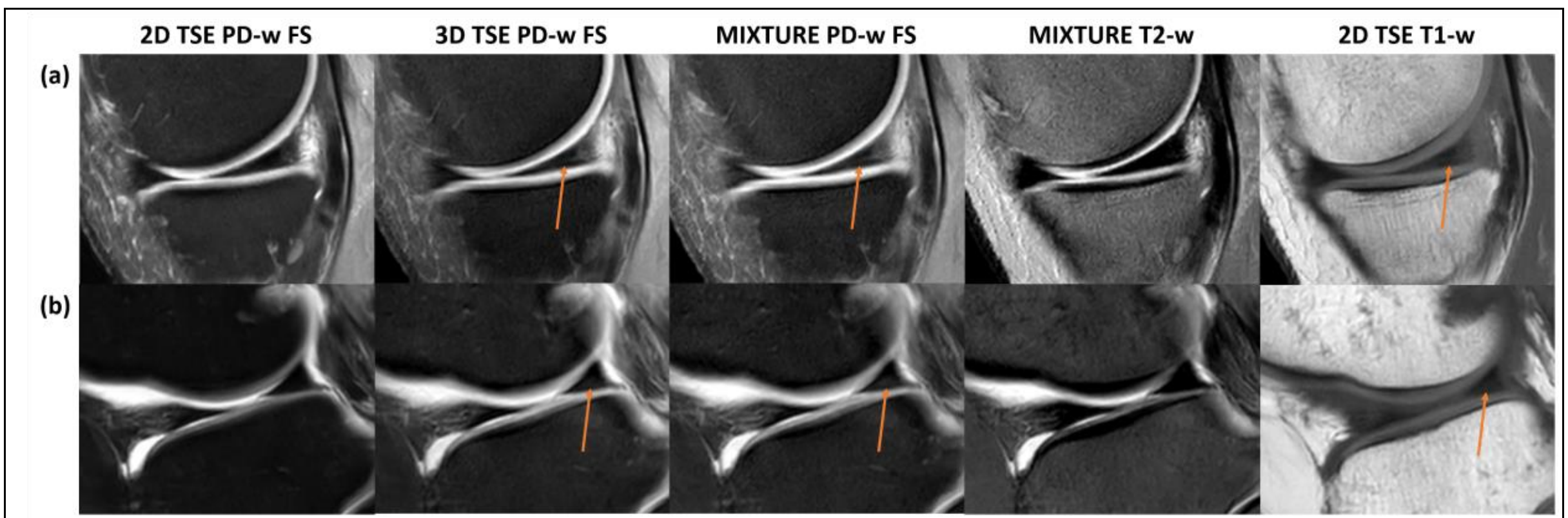

**Figure 7: Artificial findings in the meniscal posterior horn.**
Fine linear structures of high signal intensity and variable size in the posterior horns of the medial (a) and lateral (b) meniscus (orange arrows) were only appreciable on the 3D TSE and MIXTURE PD-w FS images but not on the corresponding 2D TSE or MIXTURE T2-w images. These findings are likely artificial because of the missing correlate on the T2-w image. The hyperintense structural correlate on the T1-w images suggests that the 3D TSE and MIXTURE PD-w FS images depict residual T1 contrast, which can be due to the shorter relaxation times and the utilization of complex refocusing patterns compared to the 2D TSE PD-w FS sequence. The latter had a longer TR, increasing the likelihood of complete T1 relaxation.

# Tables

## Table 1

**Table 1**: **MRI sequence parameters.** MIXTURE sequences combine morphologic imaging with quantitative parameter mapping. The sequences were acquired as PD-w FS images with quantitative T2 maps (“MIX 1“) and as T1-w images with T1ρ maps (“MIX 2“). Corresponding 2D TSE and 3D TSE reference sequences of the same image weightings were acquired, too.

Abbreviations: MIXTURE - **M**ulti-**I**nterleaved **X**-prepared **T**urbo-Spin Echo with Int**U**itive **Re**laxometry, PD - proton density, FS - fat-saturated, TSE – turbo spin echo, TR – repetition time, TE – echo time, SENSE – sensitivity encoding, NSA – number of signal averages, SPAIR – spectral attenuated inversion recovery, SPIR – spectral presaturation with inversion recovery, N/A – not applicable, SL – spin lock, TSL – spin lock time, FOV – field of view.

For 3D TSE sequences, $TE_{eff}$ and $TE_{equiv}$ denote the effective and equivalent TE as mediated by the choice of the refocusing pattern. For the 2D TSE sequence that uses a constant refocusing flip angle, TE can be described by a single value.

| **Sequence Parameter** | **MIX 1** | **2D TSE PD-w FS** | **3D TSE PD-w FS** | **MIX 2** | **2D TSE T1-w** | **3D TSE T1-w** |
|---|---|---|---|---|---|---|
| Sequence type | 3D TSE | 2D TSE | 3D TSE | 3D TSE | 2D TSE | 3D TSE |
| Orientation | sagittal | | | | | |
| TR [ms] | 1200 | 3000 | 1100 | 600 | 582 | 400 |
| TE [ms] | N/A | 40 | N/A | N/A | 15 | N/A |
| $TE_{eff}$ [ms] | 125 | N/A | 125 | 22 | N/A | 36 |
| $TE_{equiv}$ [ms] | 46 | N/A | 46 | 13 | N/A | 21 |
| Echo train length [n] | 35 | 11 | 35 | 12 | 5 | 8 |
| Refocusing pattern | ‘MSK PD FS’ | ‘no’ | ‘MSK PD FS’ | ‘Spine View T1’ | ‘constant’ (110°) | ‘MSK T1’ |
| Compressed SENSE factor | 4.5 | 2.5 | 3.5 | 6 | 2 | 6 |
| NSA [n] | 1 | 2 | 1 | 1 | 2 | 2 |
| Fat saturation [1st block – 2nd block] | SPAIR – none | SPIR | SPAIR | none – SPAIR | none | none |
| T2-preparation module TE [ms] | 0 & 50 | N/A | N/A | N/A | N/A | N/A |
| SL-preparation module TSL [ms] | N/A | N/A | N/A | 0 & 25 & 50 | N/A | N/A |
| SL frequency [Hz] | N/A | N/A | N/A | 500 | N/A | N/A |
| Scan time [min:s] | 4:59 | 4:06 | 2:57 | 6:38 | 4:23 | 4:22 |
| FOV [$mm^2$] | 140 x 140 | | | | | |

| Acquisition matrix [px] | 304 × 304 | | | | | |
| --- | --- | --- | --- | --- | --- | --- |
| Reconstruction matrix [px] | 512 × 512 | | | | | |
| Fat shift direction | anteroposterior | | | | | |
| Phase Oversampling [%] | 12+12 | 30+30 | 12+12 | 12+12 | 33+33 | 12+12 |
| Slices [n] | 43 | | | | | |
| Slice thickness [mm] | 3 | | | | | |
| Slice oversampling [%] | 12 | N/A | 12 | 100 | N/A | 12 |

## Table 2

**Table 2: Radiologic evaluation of image quality as a function of MR sequence and anatomic structure.**

The diagnostic evaluability scores for individual anatomic structures and the global diagnostic quality score are indicated as a function of image contrast and sequence (organized column-wise). Data are presented as median [lower quartile; upper quartile] of the readings by three radiologists who assigned scores for each specimen (n=10). Please refer to **Table 1** for sequence details. Additional abbreviations: MM – medial meniscus, LM – lateral meniscus.

| **Scores** | | **PD-w FS** | | | **T1-w** | | |
|---|---|---|---|---|---|---|---|
| | | **MIXTURE** | **2D TSE** | **3D TSE** | **MIXTURE** | **2D TSE** | **3D TSE** |
| **Diagnostic evaluability scores** | **MM posterior horn** | 5 [5; 5] | 5 [5; 5] | 5 [5; 5] | 3 [3; 3] | 3 [2; 3] | 3 [2.25; 3] |
| | **MM body region** | 5 [5; 5] | 5 [5; 5] | 5 [4; 5] | 3 [2.25; 3] | 3 [2; 3] | 3 [2; 3] |
| | **MM anterior horn** | 5 [5; 5] | 5 [5; 5] | 5 [5; 5] | 3 [3; 3] | 3 [3; 3] | 3 [3; 3] |
| | **LM posterior horn** | 5 [5; 5] | 5 [5; 5] | 5 [5; 5] | 3 [2; 3] | 3 [2; 3] | 3 [2; 3] |
| | **LM body region** | 5 [5; 5] | 5 [4; 5] | 5 [4; 5] | 3 [2; 3] | 3 [2; 3] | 3 [2; 3] |
| | **LM anterior horn** | 5 [5; 5] | 5 [5; 5] | 5 [5; 5] | 3 [2.25; 3] | 3 [2; 3] | 3 [2; 3] |
| | **Anterior cruciate ligament** | 5 [4.25; 5] | 5 [4; 5] | 4 [4; 4] | 3 [2; 3] | 3 [2; 3] | 3 [2; 3] |
| | **Posterior cruciate ligament** | 5 [5; 5] | 5 [4; 5] | 4.5 [4; 4.5] | 3 [3; 3] | 3 [3; 3] | 3 [3; 3] |
| | **Medial collateral ligament** | 2 [1.25; 2] | 2 [1.25; 2] | 2 [1.25; 2] | 1 [1; 1] | 1 [1; 1] | 1 [1; 1] |
| | **Lateral collateral ligament** | 3 [2; 3] | 3 [2; 3] | 3 [2; 3] | 2.5 [1; 2.5] | 2.5 [1.25; 2.5] | 3 [1.25; 3] |
| | **Medial femoral cartilage** | 5 [5; 5] | 5 [5; 5] | 5 [5; 5] | 3 [3; 3] | 3 [3; 3] | 3 [3; 3] |
| | **Lateral femoral cartilage** | 5 [5; 5] | 5 [5; 5] | 5 [5; 5] | 3 [3; 3] | 3 [3; 3] | 3 [3; 3] |
| | **Medial tibial cartilage** | 5 [5; 5] | 5 [5; 5] | 5 [5; 5] | 3 [3; 3] | 3 [3; 3] | 3 [3; 3] |
| | **Lateral tibial cartilage** | 5 [5; 5] | 5 [5; 5] | 5 [5; 5] | 3 [3; 3] | 3 [3; 3] | 3 [3; 3] |
| | **Trochlear cartilage** | 5 [5; 5] | 5 [5; 5] | 5 [5; 5] | 3 [3; 3] | 3 [3; 3] | 3 [3; 3] |
| | **Retropatellar cartilage** | 5 [5; 5] | 5 [5; 5] | 5 [5; 5] | 3 [3; 3] | 3 [3; 3] | 3 [3; 3] |
| | **Extensor mechanism** | 5 [5; 5] | 5 [5; 5] | 5 [4.25; 5] | 4 [3; 4] | 3.5 [3; 3.5] | 3.5 [3; 3.5] |
| | **Femoral bone marrow** | 4.5 [4; 4.5] | 4 [4; 4] | 4 [4; 4] | 5 [5; 5] | 5 [4; 5] | 5 [4; 5] |
| | **Tibial bone marrow** | 4.5 [4; 4.5] | 4 [4; 4] | 4 [4; 4] | 5 [5; 5] | 5 [4; 5] | 5 [4; 5] |
| | **Patellar bone marrow** | 4.5 [4; 4.5] | 4 [4; 4] | 4 [4; 4] | 5 [5; 5] | 5 [4; 5] | 5 [4; 5] |
| | **Synovium** | 5 [4; 5] | 5 [4; 5] | 5 [4; 5] | 3 [3; 3] | 3 [2; 3] | 3 [2; 3] |
| | **Effusion** | 5 [5; 5] | 5 [5; 5] | 5 [5; 5] | 4 [3.25; 4] | 3 [3; 3] | 3 [3; 3] |
| **Global diagnostic quality score** | | 5 [5; 5] | 5 [5; 5] | 5 [4; 5] | 5 [4.25; 5] | 5 [5; 5] | 5 [4; 5] |

## Table 3

**Table 3 – Post-hoc analysis of the diagnostic scores.**

Based on the Cumulative Link Mixed Models, the diagnostic evaluability and overall diagnostic quality scores were compared in a pair-wise manner. The Tukey method was employed to correct for multiple comparisons by adjusting the significance levels of the pair-wise comparisons. Significant differences against the alpha-threshold of 0.01 are marked in **bold type**.

| Image Weighting | Score | 2D TSE vs. 3D TSE | 2D TSE vs. MIXTURE | 3D TSE vs. MIXTURE |
|---|---|---|---|---|
| PD-w FS images | Diagnostic evaluability (per structure) | P=0.019 | P=0.055 | **P<0.001** |
| | Global diagnostic quality | P=0.010 | P=0.294 | P=0.025 |
| T1-w images | Diagnostic evaluability (per structure) | P=0.999 | **P<0.001** | **P<0.001** |
| | Global diagnostic quality | P=0.056 | P=0.313 | P=0.565 |

## Acknowledgements

The authors would like to acknowledge the support of David Haas and Sarah Nüsser from the Institute of Molecular and Cellular Anatomy, RWTH Aachen University.